\documentclass[12pt,preprint]{aastex}

\slugcomment{Draft: To be submitted to {\it{The Astrophysical Journal}}}
\shortauthors{Zaritsky \& Gonzalez}
\shorttitle{Cluster Strong Lensing at $0.5 \le z \lt 0.9$}

\begin{document}
\title{On the Incidence of  Strong Gravitational Lensing by Clusters in the
Las Campanas Distant Cluster Survey}
  
\author{Dennis Zaritsky}
   
\affil{Steward Observatory, University of Arizona, 933
   North Cherry Avenue, Tucson, AZ 85721, USA}

\author{and}
\author{Anthony H. Gonzalez}
\affil{Department of Astronomy, University of Florida, 211 Space Sciences Bldg., Gainesville, Fl,
32611-2055, USA}

\begin{abstract} 
The observed
incidence of strongly lensing clusters exceeds
the predictions of a $\Lambda$CDM model by about
a factor of 10.
We revisit the observational side of this discrepancy by measuring the
incidence of strong lensing in a subsample of clusters drawn from the
Las Campanas Distant Cluster Survey (LCDCS). Among clusters with
$0.5 \le z \le 0.7$, the redshift range in which we focus our
search, we find two strongly lensed systems within an effective
search area of 69 sq. deg.  There is at least one other strongly
lensed systems in the LCDCS outside of this redshift range, where we
are less complete. Over all redshifts, the 
$\Lambda$CDM model produces one large arc every 146 sq. degrees. 
Assuming Poisson statistics, the probability of finding 3 or more
strongly lensing clusters in 69 sq. degrees is
0.012. 
The lensing incidence within the LCDCS is in agreement
with that derived from an X-ray selected sample
and what has been preliminarily presented from an independent optical
cluster survey. The origin of the disagreement between theory
and observations, which remains at least at the order of magnitude
scale for the $\Lambda$CDM model, lies either in the concordance
cosmological model, in the characteristics of the resulting cluster
potentials, or in the adopted source population.

\end{abstract}

\section{Introduction}

The ``cosmic concordance" cosmological model, $\Omega_m = 0.3$,
$\Omega_\Lambda =0.7$, has quickly become the standard as a result of
new observations (Type I supernovae by \cite{r98} and \cite{perl}; cosmic
microwave background anisotropies by \cite{deB00}, \cite{h20}, and
\cite{p02}) and its ability to reconcile long-standing problems (for
examples see \cite{ost}).  The only prominent partial holdout in this
avalanche of evidence had been the distribution of QSO lens image
separations (see \cite{falco}), but even that discrepancy is now
thought to be resolved \citep{keeton}. One less well known, but perhaps
more striking discrepancy lies with the incidence of giant
gravitational arcs in galaxy clusters.  \cite{bart98} found that 
the the currently favored $\Lambda$CDM underpredicts 
the strong lensing incidence among a sample of X-ray selected
clusters by an order of magnitude. Because this discrepancy is one of the few
pieces of observational evidence against what has become the accepted
cosmological model, it warrants further examination on both the
theoretical and observational fronts.  We measure the incidence of
lensing among clusters drawn from the Las Campanas Distant Cluster
Survey \citep{gon01} and compare it to other measurements and to
the existing theoretical expectations.

The potential use of the galaxy cluster mass function to probe the
characteristics and evolution of the largest gravitationally bound
structures, and therefore constrain the cosmological model,
has been theoretically demonstrated over the past decade
(see for examples \cite{fwed}, \cite{eke96}, and
\cite{evrard}). Unfortunately, there are various reasons why the study
of clusters has yet to provide a wholly satisfying constraint.
One of the most evident weaknesses had been the
lack of large numbers of known clusters at higher ($z > 0.5$)
redshifts.  This weakness is being addressed with the
many-fold growth in the number of known high redshift, massive clusters
due in large part to a variety of both published (\cite{post96},
\cite{sco99}, \cite{gon01}) and ongoing (\cite{glad00},
\cite{willick}) ground-based, optical surveys. 

A second more pernicious difficulty involves the measurement of
the cluster mass. The measurement is fraught with possible systematic
errors and is observationally time-intensive.  There are four ways to
confirm that a cluster is massive: 1) compute the
velocity dispersion using a large number
($>30$) of member galaxy redshifts (for high-z clusters this requires
several hours of observation per spectroscopic mask on 10m-class
telescopes; see \cite{lub02}), 2) measure the temperature of the
intergalactic medium using X-ray observations (this requires exposures of tens of
ksec per cluster on X-ray satellite telescopes; see \cite{je01}), 3)
construct a Sunyaev-Zel'dovich decrement map (this requires
$\sim$ 40 hours of on-source integration; see \cite{joy01}) or 4)
measure the distortion of background galaxy images introduced by the
cluster's gravitational potential (weak lensing maps require several
hours of imaging on 10-m class telescopes; see \cite{clowe}). The demanding
observational
requirements explain why there are relatively few high-z clusters with measured
masses and create a significant impediment to taking full advantage
of the hundreds of potential massive high-z clusters presented in the
new catalogs. 

A special case of method (4) is what is of interest here.
Strong lensing, where the cluster potential
distorts the image of a background image sufficiently that it appears
as an arc (\cite{lynds}, \cite{soucail}), is a rare phenomenon that
increases in likelihood, in part, as the cluster mass increases (\cite{lup99},
\cite{dal02}). The observational advantages of using strong lensing
include that a relatively modest investment of telescope time per
cluster is necessary, that it highlights the clusters with the
greatest relaxed mass, and that it is a relatively unambiguous
phenomenon at a properly selected magnitude limit. 
Theoretical
studies (\cite{bart98}, \cite{m00}, \cite{m01}) have found that the
predicted numbers of strongly lensing clusters vary by orders of
magnitude among popular cosmological models.  The principal
disadvantage of the strong lensing approach is that only a small fraction of
clusters will produce gravitational arcs, so that even the results
from large samples of clusters may suffer from small number
statistics. Nevertheless, the differences among various models are
sufficiently large that certain models can be excluded by a small
number of lensing clusters. In particular, \cite{bart98} found that
$\Lambda$CDM models produce an order of magnitude too few giant arcs
to match the observations available at that time.

We describe a study that is analogous to that conducted by
\cite{lup99}, except that our sample is drawn from an optical survey
and concentrates on a more limited redshift range ($0.5 \le z \le
0.7$). The cluster samples are entirely independent.  The typical
concerns regarding the viability and fidelity of optical cluster
catalogs (contamination and projection effects) are less relevant here
than in a measurement of the cluster mass function because any
observed strong lensing confirms the cluster. As long as one realizes
that this study provides a lower-limit on the lensing incidence, we
avoid the concerns regarding optical cluster catalogs.  In \S2
we describe our sample and the observations. In \S3 we present our
newly discovered lensing clusters, calculate the incidence of lensing
at these redshifts, compare that incidence to that found
in other studies based on independent X-ray and optical catalogs, and 
compare our results to theoretical predictions. We summarize our findings in 
\S4.

\section{The Data}

\subsection{Sample Selection}

The target clusters are a subsample of the Las Campanas Distant
Cluster Survey (LCDCS; \cite{gon01}). The catalog resulting from that
survey contains 1073 candidate clusters at estimated redshifts,
$z_{est}$, $>$ 0.35 in a statistically complete sample. For each
candidate we provide the coordinates, an estimated redshift, which is
based on the luminosity of the brightest cluster galaxy (BCG), and a
central surface brightness measurement, $\Sigma$, which correlates
broadly with standard mass estimators such as the X-ray temperature
and luminosity, and velocity dispersion. Extensive testing, in the
original catalog paper \citep{gon01}, in work using our survey data to
examine the evolution of the cluster galaxies and the BCG
(\cite{nel01a} and \cite{nel01b}), and in the preliminary stages
\citep{gon02} of the ESO Distant Cluster Survey (EDisCS) all confirm
the contamination rates ($\sim$ 30\%) and redshift uncertainties
($\Delta z < 0.1$ for $> 80$\% of the candidates) derived in the
survey paper.

Clusters in the LCDCS are detected as regions of excess surface
brightness relative to the extragalactic background, by first removing
all resolved stars and galaxies from the survey images and then
convolving the data with a 10$^{\prime\prime}$ exponential smoothing
kernel \citep{gon01}. Positive surface brightness fluctuations are
induced by light from fainter, undetected cluster galaxies and
extended emission from the halo of the brightest cluster galaxy. The
peak amplitude of the detected surface brightness, $\Sigma$,
effectively measures the integrated cluster light within a fixed,
exponentially weighted aperture after removal of the brightest
galaxies in the cluster. The value of $\Sigma$ corrected for Galactic
extinction is $\Sigma_{cor}$.  Because of the shallow nature of the
exposures (exposure times $\sim 190$ sec on a 1m telescope), only a
few ($< 10$) cluster galaxies are typically detected and removed
before smoothing (including any resolved interlopers).  $\Sigma$ is
tied to the total cluster luminosity, with the caveats that more
galaxies will be removed prior to smoothing in nearer systems, that
the BCG may contribute significantly, and that the concentration of
the cluster will affect $\Sigma$.  These distance-dependent effects
must be empirically calibrated and that calibration is currently
somewhat uncertain due to the small number of X-ray clusters at $z >
0.7$ that we have been able to observe with the same instrumental
configuration.
For more details on this calibration we refer the reader to \cite{gon01}.

Using $\Sigma_{cor}$, we select the clusters that are likely to be the
most massive systems in the catalog at $z_{est} \ge 0.5$. Because of 
the possible exclusion of massive clusters due to
the scatter between $\Sigma_{cor}$ and mass, the measured incidence of
lensing from this sample, if quantified as the number of lenses per
unit volume (or per steradian), will be a lower limit on the true
incidence. The two other potential problems with the LCDCS selection,
the possible presence of ``false'' clusters in the sample due to
contamination and the possible inclusion of less massive clusters because
of scatter in $\Sigma_{cor}-$mass relation,
will make our search less efficient but will not bias the
lensing incidence if we express the quantity in terms of lenses per
steradian. However, potential unknown
biases, for example if the success of our method in identifying a
cluster depends on a cluster property that also affects the likelihood
of cluster lensing, such as the concentration of the mass profile,
then some possible choices for the measure of the lensing incidence, 
such as the
{\sl fraction} of lensing clusters would be artificially inflated,
while others, such as the number of lenses per steradian would be
unchanged. Care must be taken in defining a quantification of
incidence that is consistently a lower limit. We shall quote lensing
incidence as a function of area of sky.

From the LCDCS catalog, we select candidate clusters with $0.5 \le
z_{est} \le 0.85$ and $\Sigma_{cor} > 9.03 \times 10^{-3}$ cts s$^{-1}$
sq. arcsec$^{-1}$, which corresponds to a
velocity dispersion, $\sigma$, greater than 
390 km s$^{-1}$ at $z = 0.5$, 765 km
s$^{-1}$ at $z = 0.7$, and 1040 km s$^{-1}$ at $z = 0.8$ 
\citep{gon01}.  These criteria
produce a list of 125 candidates that we then classify visually. Some
high surface brightness candidates are obvious low surface brightness
galaxies or tidally interacting pairs and can be rejected immediately
(of the 125 candidates, only six were identified to be such from the
original survey data). Of the remaining 119 candidates, we select 66
as being the most promising, with the highest priority for
observation.  Independently selected, there is a sample of 30 LCDCS
candidates that are being observed as part of the ESO Distant Cluster
Survey, or EDisCS (19 of which fall within our selection criteria). We
did not re-observe the 16 of these that fall within our high-priority
list, nor the three that fall within our low-priority list.

\subsection{Observations}

We were able to observe 44 candidate clusters (40 from the high
priority list and four from the low priority list) with the Baade 6.5m
telescope on Las Campanas using the Magellan Instant Camera (MagIC)
during 9 - 12 Mar 2002. The camera's detector is a SITe 2K by 2K
device with a pixel scale of 0.069 arcsec pixel$^{-1}$ (field-of-view
is 2.36$^\prime$).  We observed in the SDSS $r^\prime$ filter
\citep{sloanfil}, with complementary $g^\prime$ observations for four
clusters that showed interesting features in the initial $r^\prime$
frames. The exposure times are either 1800 or 2400 sec total, split
between 3 frames (the total exposure times for each cluster are given
in Table 1). The seeing was excellent for the four night observing
run. The effective seeing (after registering the images and combining)
is typically 0.5 to 0.6 arcsec, and is given for each cluster in
Table 1.

The data reduction is only slightly more complicated than usual because the camera uses
a four amplifier read-out mode to reduce the readout time. Therefore, bias and flatfielding
is done differently for each quadrant. We applied the standard, public MagIC
reduction pipeline to do the first order corrections and corrected for cosmic
rays using the COSMICRAYS task in IRAF\footnote{IRAF  is the Image Reduction
and Analysis Facility written by the NOAO, which is operated by AURA under
cooperative agreement with the National Science Foundation.}.
We then use the Source Extractor (SExtractor) algorithm \citep{sx} to
produce a background sky image (objects removed). We remove the
spatially-variable background by subtracting this image from the
original image.  Images are then registered and combined using IMALIGN
and IMCOMBINE tasks in IRAF. We run SExtractor on the final combined
images to produce a catalog of objects in each target field and
objects with stellarity index $<$ 0.9 are considered to be
galaxies. Our detection parameters are set conservatively to 
require a minimum of 10 pixels, each above a 1$\sigma$ deviation above
the global sky, for an object detection. 
We interactively define polygon apertures in the $r^\prime$
band to measure the magnitudes and colors of the lensed galaxies.
Photometric calibration was done using both the SDSS calibration
\citep{sloanfil} and Landolt (1992) standards to place the photometry
on both systems.

\subsection{Sample Properties}

The redshift and surface brightness, $\Sigma_{cor}$, distributions of
the entire sample of 125 candidates and the subsample of observed
clusters (including the EDisCS clusters) are shown in Figure
\ref{fig:sampleprop}.  The only qualitative difference 
between the low and high priority clusters is that the
high priority clusters are slightly more weighted toward lower
redshifts. This difference is the result of lower-redshift candidates
appearing more secure in the original survey data, and hence assigned a higher
priority. The EDisCS observations will help fill in the high redshift
tail of the range, while most of the clusters observed here lie
between $0.5 < z_{est} < 0.7$.  The right panels of the Figure
illustrate the difficulty in defining a mass-limited sample. The steep
rise in the number of clusters toward smaller masses means that a
small systematic error could easily change the sample size by a factor
of two and that random errors will generate significant contamination
of the sample by lower-mass clusters.

The final desired quantity, the incidence of giant arcs, can be made
independent of the uncertainty in the mass measurement by being
sufficiently conservative in the mass cut used to define the imaging
subsample. Because only high-mass clusters produce strong
arcs (see \cite{lup99}), a sufficiently low mass cut will ensure that
all strongly-lensing clusters are in the subsample.  The
disadvantage of such an approach is that the number of clusters 
to observe is correspondingly larger.  The advantage of a
catalog with more precise mass measurements, such as X-ray luminosity
or temperature, is primarily in the efficiency with which one can
initially identify the candidate lensing clusters, but all studies
should find the lensing clusters.

To examine the effects of our adopted mass cut, we plot $\Sigma_{cor}$
vs. $z_{est}$ for our sample and two curves that illustrate where our
calibration of the $L_X-\Sigma_{cor}$ relationship places the two
$L_X$ limits 
from previous studies\footnote{We convert all luminosities from the literature
for an $\Omega_m = 1$ cosmology 
to the open cosmology ($\Omega_m = 0.3, \Omega_\Lambda = 0$)
used by \cite{gon00} in determining the
various relationships used here. For this reason, the numerical 
values are generally $\sim$ 1.2 times the published values.} 
in Figure \ref{fig:masscut}.
The lower curve corresponds to $L_X = 4.7 
\times 10^{44} h_{50}^{-2}$ ergs s$^{-1}$,
the limit below which previous studies found no strongly
lensing clusters \citep{lup99}. The upper curve corresponds to $L_X =
1.2 \times 10^{45} h_{50}^{-2}$ ergs s$^{-1}$, 
a cluster luminosity at which previous studies
found that over 50\% of clusters act as strong lenses
(\cite{lup99}, \cite{dal02}).  Our sample
would include all of the most-likely LCDCS lensing clusters ($L_X > 1.2 \times 10^{45}  h_{50}^{-2}$ ergs s$^{-1}$)
for $0.5 \le
z \le 0.7$, {\sl if there was no scatter in our $L_X - \Sigma_{cor}$
calibration}.  We include clusters below the $L_X = 4.7 \times 10^{44}
 h_{50}^{-2}$
ergs s$^{-1}$ line at low $z$ because our estimation of $L_X$ using
$\Sigma_{cor}$ has roughly a factor of two scatter \citep{gon01}. At
$z \sim 0.7$ the scatter in the relationship implies that we are
potentially missing some strongly-lensing clusters in our survey.
An important aspect of the calibration that is yet not well-measured is the
redshift evolution. In making the Figure we adopt a $(1+z)^4$
evolution in the optical surface brightness, even though a fit to
limited data available suggests a steeper slope \citep{gon00}. Adopting
the latter behavior would make the curves decline more steeply than
shown in the Figure and would suggest that we are even less complete
in our sample of high-mass clusters at high-redshift than illustrated
in Figure \ref{fig:masscut}.  We are currently unable to be
sufficiently conservative in our inclusion of all possible lensing
clusters at these redshifts because of the limited number of
candidates that we can follow-up. As such, this adds another reason
why we consider our result to be a lower limit on the incidence of
lensing at these redshifts.

\subsection{Arcs}

We visually examine each of the cluster images.  To match previous
work we are attempting primarily 
to identify arcs that extend at least $\sim$ 10$^{\prime\prime}$
and have $R<21.5$. These criteria place these images well above
our detection thresholds.
In the last column
of Table 1, we present comments regarding the candidate cluster
(`marginal' reflects the status of the cluster candidate, not of any
potential arc-like feature). We identify three clusters that
exhibit features that appear to be gravitationally lensed galaxies
(LCDCS 280, 486, and 954, see Figure 3-5).  For these three clusters
we also identify several additional (sometimes questionable) features
that may be lensed images. However, we do not identify similar low-quality
features in other cluster fields because of their uncertainty as
lensed images. A summary of their magnitudes and colors is presented in 
Table 2.
The magnitudes of the ``lensed" images ranges from 20.07 to 24.25 in
$r^{\prime}$, while the faintest detected objects in the frame are
typically at least two magnitudes fainter. The galaxy count histograms
typically peak at a magnitude of between 23 and 24, and only thereafter
exhibit serious incompleteness. No cluster has a histogram that peaks 
brighter than $r^{\prime} = 22$, which is 2 magnitudes fainter than our 
criteria for identifying the principal lensed image in a cluster.

\subsubsection{LCDCS 280}

This is the least convincing case of lensing among the three clusters
in which we have found ``lens-like" images. Both lens features 1 and 2
consist of four peaks, each around a different apparent mass center in
the cluster. The luminous galaxy near lens feature 1 is a typical,
extended brightest cluster galaxy.  Lens features 3,4,5, and 6 are all
quite marginal cases of lensing, but they appear curved with a center
of curvature near the BCG.  Spectroscopy of lensed features 1 and 2 is
possible with a large telescope (sources are brighter than $R = 22$
and the irregular structure may be indicative of emission line
regions). 

Assuming that the images are lensed,
we estimate the mass producing the lensing by adopting the
singular isothermal sphere (SIS) potential, measuring the distance from
the lensed image to the center of curvature ($R_L$), and making an assumption 
that the source distance is twice that of the lens. As such, the
velocity dispersion of the SIS potential is given by $\sigma_{300} =
\sqrt{(R_L/1.3^{\prime\prime})}$.  Defining the center of
curvature is somewhat ambiguous in these cases, but we explore two possibilities:
1) we visually fit a circle through the suspected lensed images 
($R_L = 7.5^{\prime\prime}$), and
2) we take the center to be coincident on the nearby dominant galaxy 
($R_L = 3.6^{\prime\prime}$).
In the
former case, we calculate that $\sigma = 720$ km s$^{-1}$, and in the latter that $\sigma = 
500$ km s$^{-1}$. The latter case suggests that if these are lensed
images, we may be seeing group/galaxy lensing rather than cluster
lensing. The former is a plausible cluster lens velocity dispersion, but 
is significantly larger than the dispersion calculated from $\Sigma_{cor}$,
520 km s$^{-1}$, using
the relationship between $\Sigma_{cor}$ and $\sigma$ from \cite{gon01}.

Given the various difficulties in convincingly identifying these
images as lensed images and our intent to maintain our measurement as
a lower limit, {\sl we will not count this object as a lensing
cluster}. Further observations are necessary to confirm the lensed
nature of these images.

\subsubsection{LCDCS 486}

This cluster contains the brightest of the arc features found in this
study (and the brightest of any of the arcs listed in the Luppino et
al. (1999) compendium; their Table 2). We measure that the arc has 
a length, $l$, of 13.7$^{\prime\prime}$ a length-to-width ratio, 
$l/w$, of 10.9 (12.5 if the width has the effect of seeing removed
by simple quadrature), and an average surface brightness of $R = 22.9$
mag sq. arcsec. 
The one potentially suspicious aspect of
this arc is the sharp bend in the middle. A possible interpretation is
that we are seeing a superposition of an arc (or two) and an unlensed
galaxy.  However, the $g^{\prime}-r^{\prime}$ color is highly uniform
along the entire length (the color of the entire arc matches that of
the primarily horizontal portion, 1a, and that of the primarily
vertical portion, 1b, to 0.02 mag). The rather unusual arc morphology
implies a challenge to the modeling of this system, but the length,
the degree of curvature, and uniform colors suggest that this is
indeed a lensed image.

Applying the same simple mass model as above (for
a measured $R_L = 9^{\prime\prime}$,
we estimate the velocity
dispersion of the corresponding SIS model to be 790 km s$^{-1}$.  From 
$\Sigma_{cor}$ we
estimate that $\sigma = 750$ km s$^{-1}$. 
The agreement between the two $\sigma$ estimates 
further supports the lensing interpretation of this image.

\subsubsection{LCDCS 954}

This is the most ``classic" lens of the three systems, 
with some other lensed features also visible in the image. In particular,
lens feature 2 is a potential radial arc. The principal arc
has $l  = 9.4^{\prime\prime}$, $l/w = 9.0$ (11.7 when the width is
corrected for the effect of seeing), and an average surface
brightness of $R = 23.8$ mag/sq. arcsec. This is the most
massive (as measured by $\Sigma_{cor}$ corrected for redshift) cluster in our
sample for $0.5 \le z < 0.7$
and shows the most dramatic lensing.  Such \"uberclusters are exceedingly rare. Given the effective area of the LCDCS (69 sq. deg; \citep{gon01}) and the discovery
of one such system at these redshifts, we expect 
only about 600 similar clusters in the Universe at $0.5 < z < 0.7$. 
There is a similar
known cluster RX J1347.5$-$1145 in the LCDCS (LCDCS 829) 
at slightly lower redshifts, 0.45, 
(which also produces gravitational
arcs; \cite{sahu98}) and has a measured bolometric $L_X$ somewhere between $9 \times 10^{45}  h_{50}^{-2}$
and $2.4 \times 10^{46} h_{50}^{-2}$ergs s$^{-1}$; \cite{sc97}, \cite{et01}).
For a local comparison, the Coma cluster has
$L_X = 2.4 \times 10^{45}  h_{50}^{-2}$ ergs s$^{-1}$ \citep{eb98}.
For a distant comparison, MS1054.4$-$0321, which 
is the best-studied X-ray luminous massive 
cluster at $z \sim 0.8$ has $L_X = 5.7 \times 10^{45} h_{50}^{-2}$ \citep{je01}, with a well-measured weak 
lensing signal \citep{h00} but no strong lensing. LCDCS 829 is known to be more X-ray
luminous than these two well-known clusters, and LCDCS 954 is expected to be
more luminous based on its value of $\Sigma_{cor}$.

Again, we estimate the mass of the system by fitting an SIS model
for a measured $R_L = 21.8^{\prime\prime}$. We calculate
that $\sigma = 1230$ km s$^{-1}$. The estimated $\sigma$ derived from $\Sigma_{cor}$ is 
1670 km s$^{-1}$. There is some discrepancy in these estimates, but both suggest
that this is a very massive cluster.

\section{Discussion}

Although we have repeatedly noted that this study will provide a lower
limit on the lensing incidence, a measurement that is far below the
true incidence may provide little constraint on cosmological
models. There are two primary reasons why our measurement could be a
gross underrepresentation of the true value. First, the LCDCS could be
severely incomplete in the most massive clusters.  Second, even if the
LCDCS is complete, our subsample could be incomplete if
our conversion between $\Sigma_{cor}$ and mass is faulty. We argue
against both of these possibilities below and then present a
comparison of our measurement of the lensing incidence to
observational and theoretical studies.

\subsection{Number Density of Massive Clusters}

To determine whether the LCDCS is missing a significant fraction
of potential lensing clusters, we compare the number densities of 
massive clusters in the LCDCS and in 
the {\it Einstein Medium Sensitivity Survey} (EMSS).
We restrict our analysis to 0.4$<$$z$$<$0.6,
where the $T_X-\Sigma_{cor}$ relation is best determined (and does not
involve the highly uncertain redshift correction), and to
$L_X>10^{45}h_{50}^{-2}$ ergs s$^{-1}$ cm$^{-2}$, where the EMSS catalog
is less incomplete.

Within the effective area of the primary, statistical LCDCS catalog 
(69 sq. degrees), we detect 9 cluster candidates
at 0.4$<$$z$$\le$0.6
with surface brightnesses that imply $L_X$$>$$10^{45}h_{50}^{-2}$ ergs s$^{-1}$.
This result yields a
raw angular density of (13$\pm4\pm4)\times10^{-2}$ clusters per square degree
(to minimize dependence on cosmological models we 
calculate and compare angular densities).
The first quoted uncertainty reflects Poisson
counting; the second reflects the systematic
uncertainty from the conversion of $\Sigma_{cor}$ to $L_X$ (a 1$-$$\sigma$ variation in this
relation results in a net change in sample size of $\pm$4 clusters). 
There are several further steps in converting this raw density to a true density.

Because the number of clusters rises rapidly as mass
decreases, significant scatter in $\Sigma_{cor}$ or $L_X$, which is certainly present
in the LCDCS, will artificially boost the number of clusters found
above an imposed mass or $L_X$ threshold.
Redshift uncertainty and
contamination in the LCDCS catalog also act to amplify the observed
number of clusters.  To assess the effect of scatter in the $L_X$
conversion on the number density, 
we approximate the underlying cluster mass function with a
Press-Schechter distribution \citep{pre74}. 
This distribution is then multiplied by a transfer function to reproduce the 
effect of imposing a fixed mass limit. For data with a Gaussian uncertainty
in mass, this yields
\begin{equation}
\frac{d\tilde N(M)}{dM} = \frac{dN(M)}{dM}   \frac 
	{1-\mathrm{erf}((M-M_{lim})/\sqrt{2}\sigma_M)}{1+\mathrm{erf}(M/\sqrt{2}\sigma_M)},
\end{equation}
where $d\tilde N/dM$ is the number of clusters of a given mass that 
are included in the sample, $dN/dM$ is the initial Press-Schechter mass function,
$\sigma_M$ is the uncertainty in $M$, and $M_{lim}$ is the
mass limit of the survey imposed by the observer.  
To connect virial mass to
$L_X$, we use the $L_X-T_X$ relation of \citet{xue2000} in conjunction with the
equation \begin{equation} M_{vir}=10^{15} \left(\frac{1.15 \beta T_X}{
1+z}\right)^{3/2} [\Omega_0 \Delta_{vir}(z)]^{-1/2} h_{50}^{-1}
M_{\odot}, \end{equation} which relates virial mass to X-ray
temperature for a spherically symmetric, isothermal plasma in virial
equilibrium (cf. \citet{eke96} and \citet{bor99}).  The quantity
$\Delta_{vir}(z)$ is defined as in \citet{kit96}, and we set
$\beta$=1. For the LCDCS, the
uncertainty is consistent with arising primarily from observational
uncertainty (i.e. $\sigma_\Sigma$ is constant), 
so $\sigma_M/M\approx3.45\times\sigma_\Sigma/\Sigma\approx 0.42 		
(L_X/10^{45})^{-0.14}$ for $\sigma_{\Sigma}=1.85\times10^{-3}$ 
counts s$^{-1}$ arcsec$^{-2}$ (see \cite{gon01} 
for the origin of this value).
The EMSS has the same qualitative problem, but 
the uncertainty in $L_X$ is dominated by intrinsic scatter (i.e. the fractional
uncertainty $\Sigma_{L_X}/L_X$ is roughly constant), so
$\sigma_M/M\approx0.5\times\sigma_{L_X}/L_X\approx0.23$ for $\sigma_{\log L_X}$=0.2. For an imposed threshold of 
$L_X=10^{45} h_{50}^{-2}$ 
ergs s$^{-1}$ cm$^{-2}$ there will be $\sim$21\% more clusters 
included in the LCDCS sample than in the EMSS sample due to the larger
scatter in optically estimated mass.

Redshift uncertainties lead to a similar bias in which
the more numerous, poorer clusters at $z$$<$0.4 are scattered 
into the relevant redshift range. Although poorer, the surface brightnesses
of these systems can be comparable to the more massive clusters at $z$$>$0.4
because of the redshift dependence of $\Sigma$. 
We again use the PS formalism to compute the expected number of clusters
above the LCDCS detection threshold as a function of redshift, and then 
convolve this distribution with the redshift uncertainty (see \citealt{gon01}
for the functional form of the redshift uncertainty).
The redshift bias introduces an additional 19\% increase in the observed 
surface density.

Finally, contamination of the LCDCS 
sample by sources such as low surface brightness galaxies also boosts the 
observed surface density. For the redshift range 0.4$<$$z$$<$0.6, we 
expect, on the basis of spectroscopy and imaging, 
that $\sim$ 80\% of the sources are 
real clusters \citep{zar97, gon01, nel01a}. Accounting
for all
of the issues described above yields a corrected LCDCS surface density
of $(7.2\pm2.4\pm2.4)\times 10^{-2}$ clusters per square degree with 
$L_X$$>$10$^{45}$$h^{-2}_{50}$ ergs s$^{-1}$.

The EMSS contains five systems with bolometric luminosity
$L_X$$>$10$^{45}h_{50}^{-2}$ ergs s$^{-1}$ cm$^{-2}$ in this redshift
range. 
Following the method of \citet{hen92} and \cite{lup95}, we compute the 
number density of EMSS clusters and obtain
a comoving density of 9.6$\times10^{-8}h^{3}$ Mpc$^{-3}$
for $\Omega$=1, $\Lambda=0$, which corresponds to 
a projected density of (1.6$\pm$0.7)$\times10^{-2}$ clusters 
per square degree, where the quoted error is purely Poissonian. 
This value is 4.5 times smaller than the LCDCS value.

Why such a large discrepancy? 
Massive clusters are strongly correlated (see \cite{gon02a} for an
analysis of the LCDCS) and
the LCDCS sample includes RX J1347.5, which is the most luminous 
known X-ray cluster. Several other massive clusters are in close proximity 
to RX J1347.5, and consequently, the inadvertent inclusion of 
this particularly rich 
region of space in the LCDCS positively biases our computed angular density.
To estimate the magnitude by which this correlation
may impact the results, we recompute the angular density using 
only clusters that are at least
30 arcmin away from RX J1347.5 ($\sim 100 h^{-1}$ Mpc at z=0.5). 
With this restriction we have only 4 clusters in the remaining 60 square 
degrees (i.e. 56\% of the most massive clusters in the LCDCS at 0.4$\le$$z$$<$0.6 
are located within a region corresponding to 15\% of the total survey area).
Including correction for the statistical biases cited above 
yields a revised
angular density of (3.7$\pm1.8^{+1.8}_{-0.9})\times10^{-2}$ clusters per square degree for the LCDCS.
This value is 50\% lower than that
derived for the entire survey and not significantly ($< 2\sigma$)
discrepant with the EMSS value.
We conclude that the two surveys are marginally consistent with one another
but that even in surveys that span over 100 sq. degrees
it is possible
to be significantly (factor of two) affected by large scale clustering.
In particular, given that the number density from the LCDCS is larger 
than that from the EMSS, 
we conclude that the LCDCS is not missing a large number of potential
lensing clusters out to at least $z = 0.6$.

\subsection{Testing the $\Sigma_{cor}-$Mass Correlation}

The degree to which we have identified all of the possible lensing clusters
within the LCDCS for our follow-up imaging depends on the validity of the $\Sigma_{cor}-$mass
relationship and its scatter. 
The empirical relation between the surface brightness of the original low
surface brightness detection image and the mass of a cluster is only broadly
established \citep{gon01}. We will continue to test and refine this 
relationship because with a large survey it is necessary to develop a method to 
estimate cluster masses, even crudely, from the original survey data. Subsamples
can then be explored in greater detail, as done here. The EDisCS
program will provide velocity dispersions and weak lensing maps for
$\sim 20$ LCDCS clusters. Here, with a larger sample, but cruder data,
we examine whether the number of luminous galaxies in a cluster, which 
is correlated with cluster mass (see \cite{koch}), correlates
with $\Sigma_{cor}$.

 In Figure \ref{fig:ngal} we plot the number of galaxies within our images
brighter than a limit set above our estimated completeness,
N$_{GAL}$. We do not correct for the differences among estimated
luminosity distances for the clusters
because such corrections correlate for N$_{GAL}$
and $\Sigma_{cor}$, and hence the redshift errors would create an apparent 
correlation. We exclude the cluster candidate confirmed to be
a low surface brightness galaxy (LCDCS 801)
and the two suspected as arising from tidal debris among interacting
galaxies (LCDCS 857 and 899).
Despite the lack of a correction for the redshift effects on either
N$_{GAL}$ or $\Sigma_{cor}$, the lack of a correction for any
foreground/background galaxy contamination (our fields-of-view are too
small to do a locally-determined background correction), and the
inclusion of the marginal clusters, there is a weak correlation
between $\Sigma_{cor}$ and N$_{GAL}$ (Spearman rank correlation
coefficient 0.22, probability of random occurrence 0.15). Although the
correlation has significant scatter, of the 14 cluster fields with $>$ 80
galaxies in our images all but one have $\Sigma_{cor} \ge 10.1$, the
median $\Sigma_{cor}$ of our sample. Selecting high
$\Sigma_{cor}$ clusters does not guarantee clusters with many galaxies,
but it does nearly guarantee that such clusters will be among the
sample. The scatter in the correlation comes primarily from clusters
with high $\Sigma_{cor}$ that upon further examination appear to be
poorer clusters than expected. The high measured value of
$\Sigma_{cor}$ is presumably the result of a poorly subtracted galaxy
halo, scattered light, coincident Galactic infrared cirrus, or some
other contamination.  
Although it is evident that significant scatter exists and a sample
sharply defined by $\Sigma_{cor}$ will not be sharply defined in any
other cluster property (such as richness or mass), it is sufficient
for culling, albeit sometimes inefficiently, the richest 
(and presumably most massive) clusters.

\subsection{Comparison With Previous Surveys}

\subsubsection{Incidence of Lensing}

Accepting the two lensing clusters as the only ones in the LCDCS sample at these
redshifts, then
the incidence of clusters that produce giant arcs over the redshift range 
$0.5 \le z \le 0.7$ that are brighter than $R \sim 21.5$ and 
have $l/w \ge 10$
is $\ge 0.029 \pm 0.020$ deg$^{-2}$.  The uncertainties reflect only
Poisson noise (they do not include the effects of large scale structure
and do not reflect the various reasons discussed previously that
make this a lower limit on the lensing incidence).

\subsubsubsection{\cite{lup99}}

The largest directed search for giant arcs in clusters is the study
by \cite{lup99} of 38 X-ray selected clusters, $L_X > 2.4\times 10^{44} h_{50}^{-2}$ ergs
s$^{-1}$ at $ 0.15 \le  z \le 0.823 $. 
They find strong lensing ($l \ge 8^{\prime\prime}$ and $l/w \ge 10$) in 8
of the 38 clusters (plus some arclets and some suggestive features).
Of the most X-ray luminous clusters ($L_X > 1.2 \times 10^{45}  h_{50}^{-2}$ ergs s$^{-1}$)
they find 60\% (3 of 5) contain giant arcs, while none (0 of 15)
of the least X-ray luminous clusters ($L_X < 1.2 \times 10^{44} h_{50}^{-2}$ ergs s$^{-1}$
contain giant arcs.). These are clusters drawn from the
{\sl Einstein Observatory} Extended Medium Sensitivity Survey (EMSS;
\cite{gio90}, \cite{st91}) which for $\delta > -40^\circ$ 
covered 734.7 sq. degrees \citep{hen92}. 
Most of their clusters are at $z < 0.5$ because the X-ray survey is most
sensitive to the nearer clusters.

They detect 2 clusters with giant arcs at $z> 0.5$. To compare to the 
LCDCS we need to determine the survey area over which they would
have found these clusters. The survey is not complete to all flux limits in
the catalog across all fields.
For MS0451.6-0305 (z = 0.55, $L_X = 23.7\times 10^{44}  h_{50}^{-2}$ ergs s$^{-1}$), the effective survey area is 720.6 sq. degrees.
For MS2053.7-0449 (z = 0.583, $L_X = 6.8\times 10^{44} h_{50}^{-2}$ ergs s$^{-1}$), the effective survey area is 225.4 sq. degrees.
Naively, one would simply add 1/720.6 and 1/225.4 to obtain the lensing
incidence of 0.0058, which is smaller than the LCDCS rate. However,
part of the reason that their value is so low is because the EMSS survey
is incomplete for clusters at redshifts between 0.5 and 0.7.
For a given
$L_X$ each cluster is observed over some fraction of the volume between
these redshifts. For MS0451 that fraction corresponds to 0.93, while
for MS2053 it corresponds to 0.32. Assuming that the number density of 
lensing clusters is uniform throughout this volume, we can correct the
observed values for incompleteness. Doing so, we derive that the
lensing incidence measured using the EMSS and the \cite{lup99}
study  for $z > 0.5$ is $0.014 \pm 0.010$ deg$^{-2}$.
We conclude that our measurement
is in agreement with that from the EMSS, although both measurements
are compromised by small number statistics. Although we cannot proceed
further with this comparison due to the small number of systems involved, 
the marginally lower lensing incidence in the EMSS area at $z > 0.5$ may
in part be connected to the lower number of massive clusters identified in the EMSS
relative to the LCDCS at these redshifts (see \S3.1).

\subsubsubsection{Gladders, Yee, and Ellingson (2002)}

\cite{glad02} present the discovery of one high-redshift (0.77)
cluster with strongly-lensed arcs. However, they also provide a preview
of what their entire survey might find. This is an interesting, if
slightly
premature, comparison because their survey is based on an independent
method of finding
clusters over a comparable survey volume. They have found six
clusters with strong lensing to date in their survey, but because
\cite{glad02} focus on the one at $z = 0.77$ there are few details 
of the others (an image of another at $z = 0.62$ is presented by \cite{yee02}). 
The six have been found in $\sim 50$\% of the survey
data (the survey in its entirety will cover $\sim 100$ sq. deg).
The photometric redshifts of the lensing clusters are all $>$ 0.5. Therefore,
the complete survey may be expected to yield 12 strong lensing clusters
at $z > 0.5$. To best match the situation with the RCS, we include
one LCDCS lensing cluster (RX J1347.5) at slightly below $z = 0.5$ because
the photometric redshift uncertainty would allow it to be at $z \ge 0.5$ 
and one LCDCS cluster observed as part of the EDisCS survey \citep{white02}
that shows arcs even though its arcs may not satisfy the $l/w \ge 10$ criteria
that we will use to compare to the simulated lensing incidence.
From these four LCDCS lensing clusters we 
set a lower-limit incidence of lensing at $z > 0.5$ of 
$0.058 \pm 0.029$ deg$^{-2}$, about half of what is preliminarily 
the case for the Red Cluster Sequence survey 
($0.120 \pm 0.049$ deg$^{-2}$), but within $\sim 1.5\sigma$.

Although the discrepancy is only marginally significant, 
there are various possible explanations for a difference:
(1) the RCS lenses
may not satisfy the same criteria ($l$, $l/w$, or magnitude limit)
as the LCDCS lenses and so the numbers may not be directly comparable, (2)
there may be more lenses lurking in the LCDCS (we have not deeply
imaged all rich-cluster candidates), and (3) the LCDCS and RCS may
select different clusters, with the LCDCS in some way selecting
against ``good" lensing clusters. We cannot comment further on possibility
(1) until the full RCS results are published. 
Examining Figure \ref{fig:masscut}, and assuming that the correlation between 
$\Sigma_{cor}$ and $L_X$ is reasonably tight, we find that we 
have imaged 13 of 21 candidate clusters
expected to have $L_X > 4.7 \times 10^{44}$ ergs s$^{-1}$ within $0.5 <
z < 0.7$ (this includes low-priority targets). Making a completeness
correction suggests that one of the unobserved clusters (assuming that
the low-priority targets are indeed clusters) would have an arc.
These numbers would suggest that the solution to the possible discrepancy does
not lie entirely with unobserved clusters in our sample at $z < 0.7$. 
However, at $z > 0.7$
we see that our sample becomes much more seriously incomplete and that
because of volume effects the sample potentially includes a large
number of very massive clusters.  Among our lensing clusters at $z > 0.5$, 1 of 4 is
at $z > 0.7$ - if this ratio holds true for the RCS sample, then the full
RCS may have only 9 lensing clusters at $0.5 < z < 0.7$ and a lensing
rate of $\sim 0.090 \pm 0.030$ deg$^{-2}$, which is less 
discrepant with our measurement. The agreement in the
projected number densities of massive clusters and in the lensing
incidence between the X-ray selected EMSS and the LCDCS argues against
option (3).

Regardless of any potentially significant difference
between the LCDCS and the final RCS lensing rates, there are some key agreements
coming from these various studies: (1) massive clusters that produce
giant arcs exist out to $z \sim 0.8$, (2) lensing clusters can be
identified from optical surveys at these redshifts at the incidence of
1 to 2 per 20 sq. degrees (leading to a prediction of over 2000 such
clusters in the entire sky), and (3) the number of lenses found in
optically-selected cluster surveys is not qualitatively different than
in an X-ray selected cluster survey (although initial X-ray selection will
definitely improve the odds of starting with a sample of the likeliest
lensing clusters).

\subsubsubsection{\cite{bart98}}

The potential use of arc statistics as a cosmological constraint 
led to theoretical efforts to estimate the lensing rate in various
cosmologies. \cite{bart98} published their predictions for the 
total numbers of lenses across the sky in three archetypical models
(standard CDM (with a tilted power spectrum), open CDM, and $\Lambda$CDM).
A comparison of the predictions
and our results is of interest and provides a glimpse into what may be
possible with both more detailed simulations and larger observational
samples.

The principal result of the \cite{bart98} study is that different
cosmologies predict lensing rates that vary by orders of magnitude.
In particular, their open CDM model predicts 2400 strongly lensing
clusters and their $\Lambda$CDM model predicts 280 across the
sky. The properties of the simulated arcs 
(magnitude limit $R < 21.5$ and $l/w \ge
10$) are similar to those presented here 
(the lenses presented here are brighter than $R = 21.5$ and have
$l/w \ge 10$, $l > 8^{\prime\prime}$). 
Assuming Poisson statistics and the predicted lensing rates,
we calculate that the probability
of finding 2 or more strongly lensing clusters within 69 sq. degrees (at
any redshift) is
0.08. If we include the other known LCDCS lensing cluster that
satisfies the criteria (RX J1347.5; \cite{sahu98} identified multiple arcs, 
with one arc having $l = 7.8^{\prime\prime}$, $l/w= 15.6$, $V = 21.5$)
we find that the probability of finding 3 or more clusters within 
69 sq. degrees
is 0.012 (the brightest arc in the 
lensing cluster identified in the EDisCS survey does not satisfy 
the $l/w \ge 10$ criteria although it has $l = 11^{\prime\prime}$). 
Therefore, even the lensing incidence that we have measured, 
which is a lower limit, is in statistically significant 
contradiction to the predictions of a $\Lambda$CDM model.

To further strengthen the statistics of the argument, we combine the
results from the LCDCS and RCS. From the RCS, the images of two
lensing clusters have been published (\cite{glad02}, \cite{yee02}). 
While quantitative measures of the geometry of the arcs
are not presented, they 
appear to satisfy the $l/w \ge 10$ criteria. However, one system
is slightly fainter than $R = 21.5$ and the magnitude for the second
system is not published. Including just these two clusters, although
\cite{glad02} have identified six lensing clusters 
in roughly half the survey
area (50 sq. degrees), we have a total of five strongly lensing
clusters over 119 sq. degrees. The probability of finding this many
clusters in such an area of sky in the $\Lambda$CDM model is $1.5\times
10 ^{-3}$. Although including these systems increases the statistical
significance of the discrepancy, the discrepancy is significant 
even with only the three LCDCS systems. It is interesting to note that 
various surveys are leading to the same conclusion and that in combination
they may already provide exceedingly strong statistical constraints.

We conclude that the observations are 
discrepant with the predictions of the $\Lambda$CDM model as
constructed by \cite{bart98}. We note that the application of the
\cite{bart98} results is not entirely consistent with this
observational sample because they focused on clusters at $z \sim 0.3$
and source galaxies at $z < 1$. We are likely to have some source
galaxies at $z > 1$, although the primary lensed images are quite
bright ($R \le 21.5$) and so are unlikely to be very distant objects
(a counterexample is presented by \cite{glad02} in that one of their
arcs is a galaxy at $z=4.8786$).
As argued by Bartelmann et al., galaxies that produce arcs with $R <
21.5$ are likely to be galaxies with intrinsic $R < 23$, which are in
turn likely to be at $z < 1$.  Evidently, the models must be
constructed to more accurately represent these samples, but our
results and conclusions are in direct agreement with those of
Bartelmann et al.'s based on the results from the independent EMSS.

To increase the lensing rates in the models (see \cite{bart98}), there
have been appeals to certain cluster properties such as strong
asymmetry and presence of a massive central concentration (such as the
BCG).  In support of some missing ingredient in the models, we note
that there are now at least two known strongly lensing clusters at $z
> 0.7$, where all of the \cite{bart98} models predict a negligible
likelihood of lensing.  However, subsequent studies (\cite{m00},
\cite{fl00}) that examined the effect of substructure and asymmetries
found only moderate changes from the previous results. In contrast to
another challenge to the CDM framework, the compactness of galaxy
cores, any solutions to that challenge that affects the concentration
of all CDM cluster halos will exacerbate the discrepancy between the
predicted and observed number of strong lenses \citep{bart02}.
A recent study \citep{bartnew} finds promise in progressing
beyond the simple cosmological constant and altering the equation
of state. 

The simulations are currently challenged to achieve both sufficient
resolution to trace cluster substructure and model a sufficiently
large volume of the Universe to include enough of the most massive
clusters (for example see \cite{ham}). In particular, the scatter in
halo properties must be accounted for because the most concentrated
systems will dominate the lensing incidence \citep{wyithe}.  When such
simulations exist, one should match not only the total number of
lensing clusters, but also their redshift distribution in order to
constrain the growth of structures in the Universe and the underlying
cosmological framework. We conclude that the data are now at a
sufficient stage to motivate a new exploration of the theoretical
models.

\section{Summary}

We have examined a subsample of cluster candidates drawn from the 
Las Campanas Distant Cluster Survey (LCDCS) for evidence of strong
gravitational lensing. We conclude the following:

\noindent
1) Over the redshifts range we have explored most extensively ($0.5 < z < 0.7$)
we have identified two previously unknown strongly lensing systems and a third
possible system (see Table 1) among a subsample of LCDCS clusters. 
Using only the two secure lensing systems from these three, we set a 
lower limit on the incidence of strong lensing in this redshift range of $0.029 \pm
0.020$ deg$^{-2}$.
 
\noindent
2) Over the entire redshift range of the LCDCS ($0.35 < z \le 0.85$) 
we now know of four secure lensing systems (including one previously known one, 
RX J1347.5 and one that is part of the EDisCS sample \citep{white02}).
Of these four, the arcs in EDisCS lensing cluster may not satisfy the generally
adopted magnitude
and size criteria. Retaining only the three other lensing systems, we set 
a lower limit on the incidence of strong lensing in this redshift 
range of $0.043 \pm 0.025$ deg$^{-2}$.

\noindent
3) Our lower limit on the lensing incidence at $z > 0.5$ 
agrees within the uncertainties with
the measurement based on the results from a study of EMSS clusters 
($0.014 \pm 0.010$ deg$^{-2}$; \cite{lup99})
and the preliminary results from the Red Cluster Sequence survey 
(somewhere between 0.06 to 0.12 deg$^{-2}$; \cite{yee02}). 

\noindent
4) The empirical values for the number of strongly lensing clusters
are significantly higher than the prediction of the $\Lambda$CDM model
from \cite{bart98}.  The probability of finding 3 or more strongly lensing
clusters in the LCDCS's 69 sq. degrees is 0.012. The discrepancy between
the observed and predicted rates for the $\Lambda$CDM model is in
agreement with the conclusion of \cite{bart98} based on their
comparison to the results from the EMSS
clusters. Although our study has confirmed the discrepancy between
observations and theoretical expectations in the ``concordance" model,
we have used the same theoretical results, namely those presented by
\cite{bart98}. This discrepancy indicates a problem with the
theoretical models either with the generally accepted concordance
cosmological model (because the predictions from an
open CDM model are in much better
agreement with the observations), the specifics of the cluster formation
and resulting gravitational potential, or the assumed background
population.  The next step is to reexamine the models to see if an
order of magnitude increase in the lensing rate for what has become
the standard cosmological model, $\Omega_m =0.3, \Omega_\Lambda =
0.7$, is at all possible.

\section{Acknowledgments}

The authors thank Chuck Keeton for comments on a preliminary manuscript.
DZ acknowledges financial support from National Science Foundation
CAREER grant AST-9733111 and a fellowship from the David and Lucile
Packard Foundation.

\clearpage

\begin{figure}
\plotone{f1.eps}
\vskip 6in
\figcaption{The distribution of redshifts and central surface brightnesses of
cluster candidates. The upper panels represent the high-priority sample and
the lower panels the low-priority sample. Boxes that are shaded in only 
one direction
represent our observations. Boxes that are shaded in both directions represent
the EDisCS snapshot observations \citep{gon02}. Open boxes represent 
LCDCS clusters that satisfy the basic criteria of the sample described
in this paper, but which have not yet been imaged deeply.
\label{fig:sampleprop}}
\end{figure}
\clearpage

\begin{figure}
\plotone{f2.eps}
\vskip 6in
\figcaption{The distribution of central surface brightnesses
vs. redshift. The lower panel represents the sample presented in this
paper, while the upper represents the EDisCS snapshot sample
\citep{gon02}. In the lower panel, the filled circles represent
high-priority targets that were observed, the open circles represent
high-priority targets that were not observed, the filled squares
represent low-priority targets that were observed, and the open
squares represent low-priority targets that were not observed.  The
larger open circles indicate the three clusters in which we identified
possible gravitational arcs (the one below the solid curve is LCDCS
280, for which the nature of the lensing images is most in doubt).
The two confirmed LSB galaxies (LCDCS 587 and 801)
are not included. The dashed line corresponds to our calibration of
the $\Sigma_{cor} -L_X$ relationship \citep{gon00} for 
$L_X = 1.2 \times 10^{45} h_{50}^{-2}$ergs s$^{-1}$
where we have taken the redshift evolution to be $(1+z)^4$.
The dashed line represents the
location of $L_X = 4.7\times 10^{44} h_{50}^{-2}$ ergs s$^{-1}$ clusters.
\label{fig:masscut}} 
\end{figure}

\clearpage
\begin{figure}
\plotone{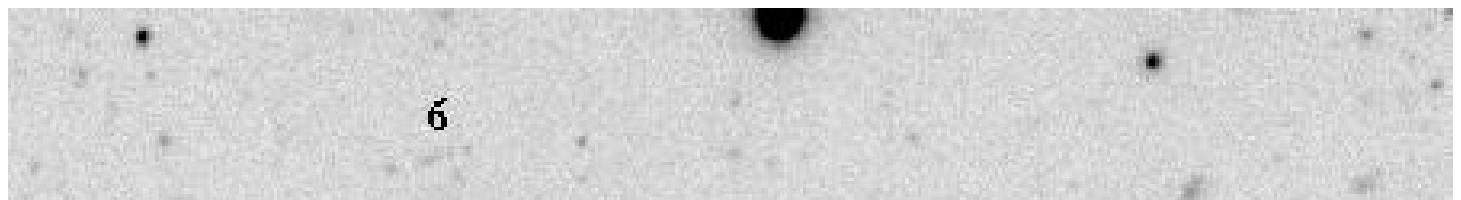}
\vskip 6in
\figcaption{The $r^\prime$ image of LCDCS 280. 
Potential arcs are labeled. 
The most likely arcs are labeled 1 and 2. The field-of-view is 
141 arcsec across, corresponding
to 896 kpc at the estimated redshift of the cluster, 0.54 (for $H_0 = 70$ km
s$^{-1}$ Mpc$^{-1}$, $\Omega_m = 0.3, \Omega_\Lambda = 0.7$).
\label{fig:f3}}
\end{figure}
\clearpage

\begin{figure} 
\plotone{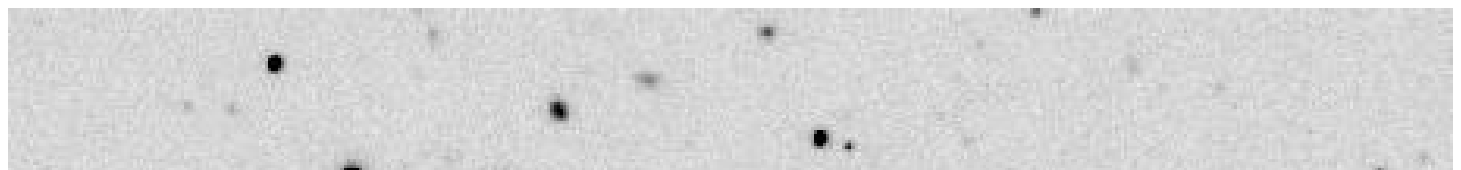} 
\vskip 6in 
\figcaption{The $r^\prime$ image of LCDCS 486. The potential arc is
labeled.  The field-of-view is 141 arcsec, corresponding to 949 kpc at
the estimated redshift of the cluster, 0.61
 (for $H_0 = 70$ km s$^{-1}$ Mpc$^{-1}$, $\Omega_m = 0.3,
\Omega_\Lambda = 0.7$).
\label{fig:f4}} 
\end{figure} 
\clearpage

\begin{figure} 
\plotone{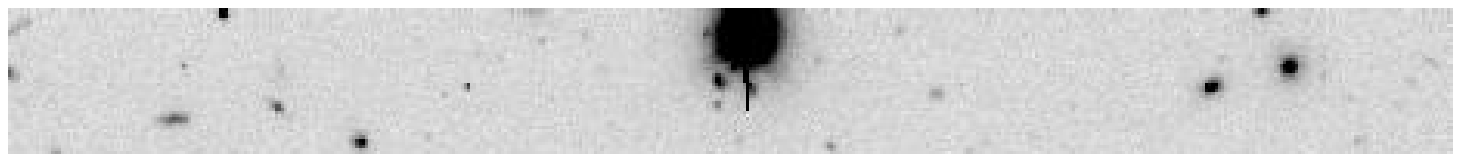} 
\vskip 6in 
\figcaption{The $r^\prime$
image of LCDCS 954.  Potential arcs are labeled.  The field-of-view is
141 arcsec, corresponding to 990 kpc at the estimated redshift of the
cluster, 0.67 (for $H_0 = 70$ km s$^{-1}$ Mpc$^{-1}$, $\Omega_m = 0.3,
\Omega_\Lambda = 0.7$).  
\label{fig:f5}} 
\end{figure} 
\clearpage

\clearpage
\begin{figure}
\plotone{f6.eps}
\vskip 6in
\figcaption{The number of galaxies brighter than $r^\prime = 26.6$ 
within the entire image is plotted 
vs. $\Sigma_{cor}$. The large open circles highlight the three
clusters with candidate giant gravitational arcs.
The dashed vertical line divides the sample into two equal subsamples
based on $\Sigma_{cor}$. The cluster near the dashed line is
LCDCS 280, which has the more dubious lensed images.
\label{fig:ngal}}
\end{figure}

\clearpage

\begin{deluxetable}{lcccccrrl}
\tablewidth{0pt}
\tablecaption{Cluster Observations}
\tiny
\tablehead{
\colhead{LCDCS} &\colhead{$\alpha$} & \colhead{$\delta$}&
\colhead{Exp.}&\colhead{Seeing}&\colhead{$z_{est}$}&  \colhead{$\Sigma_{cor}$}&N$_{GAL}$&\colhead{Comments} \\
\colhead{Number}&(2000.0)&(2000.0)&(sec.)&($^{\prime\prime}$)&&(10$^{-3}$ cts/s/sq. $^{\prime\prime}$)
}
\startdata
0005&10:02:27.1&$-$12:47:13&2400&0.53&0.57&9.17&75\\
0017&10:05:43.6&$-$11:47:43&1800&0.75&0.53&10.10&68\\
0027&10:07:54.7&$-$12:21:19&2400&0.60&0.52&10.50&93\\
0091&10:31:50.3&$-$12:44:27&2400&0.50&0.69&10.10&98\\ 
0104&10:36:10.4&$-$12:44:43&1800&0.66&0.55&9.03&55\\
0111&10:38:00.0&$-$12:18:54&1800&0.65&0.66&9.45&74\\
0163&10:51:22.7&$-$12:01:27&2400&0.65&0.67&12.30&137\\
0240&11:17:22.0&$-$11:55:46&1800&0.70&0.79&10.90&86&Marginal\\
0248&11:18:36.2&$-$12:02:03&1800&0.51&0.60&10.20&50&Marginal\\
0280&11:23:59.4&$-$11:50:07&3200&0.67&0.54&10.10&83&Lens System\\
0374&11:48:06.0&$-$11:45:33&2400&0.62&0.53&9.38&70\\
0415&11:57:34.7&$-$12:18:48&2400&0.62&0.60&10.70&73\\
0417&11:57:54.3&$-$11:51:22&1800&0.65&0.82&11.60&44&Marginal\\
0418&11:58:14.6&$-$12:14:11&1800&0.66&0.65&10.30&72\\
0468&12:10:12.7&$-$12:19:07&1800&0.68&0.74&9.42&73&Marginal\\
0486&12:12:29.9&$-$12:16:19&2400&0.62&0.61&11.10&92&Lens System\\
0550&12:35:27.7&$-$12:57:01&2400&0.47&0.57&9.12&111&\\
0586&12:44:51.7&$-$12:56:56&1800&0.64&0.58&9.36&69&\\
0589&12:45:02.0&$-$11:49:19&2400&0.51&0.50&10.50&70\\
0616&12:55:31.5&$-$12:16:55&1800&0.64&0.54&11.00&57&Marginal\\
0635&13:01:44.9&$-$12:13:24&1800&0.64&0.50&9.86&76&Marginal\\
0684&13:15:47.5&$-$11:37:26&2400&0.53&0.53&10.50&87\\
0698&13:19:50.3&$-$12:06:35&2400&0.65&0.55&18.1&132\\
0717&13:22:56.9&$-$11:44:43&2400&0.52&0.72&9.50&74\\
0719&13:23:50.2&$-$12:52:51&1800&0.48&0.59&9.17&68\\
0778&13:35:50.4&$-$11:46:16&1800&0.84&0.58&10.40&69\\
0785&13:37:08.7&$-$12:57:14&1800&0.46&0.62&14.10&51\\
0795&13:39:22.1&$-$13:00:14&2400&0.51&0.59&11.30&55\\
\enddata
\label{tab:observations}
\end{deluxetable}
\clearpage

\setcounter{table}{0}
\begin{deluxetable}{lcccccrrl}
\tablewidth{0pt}
\tablecaption{Cluster Observations (continued)}
\tiny
\tablehead{
\colhead{LCDCS} &\colhead{$\alpha$} & \colhead{$\delta$}&
\colhead{Exp.}&\colhead{Seeing}&\colhead{$z_{est}$}&  \colhead{$\Sigma_{cor}$} & N$_{GAL}$&\colhead{Comments} \\
\colhead{Number}&(2000.0)&(2000.0)&(sec.)&($^{\prime\prime}$)&&(10$^{-3}$ cts/s/sq$ ^{\prime\prime}$)
}
\startdata
0801&13:40:09.6&$-$12:10:09&1800&0.63&0.67&13.20& &LSB\\
0814&13:42:12.8&$-$12:59:28&1800&0.46&0.54&9.35&69\\
0827&13:46:42.4&$-$11:59:24&2400&0.47&0.71&11.30&81\\
0836&13:48:52.0&$-$12:04:18&1800&0.43&0.59&12.80&53\\
0857&13:54:50.9&$-$12:09:11&1800&0.45&0.72&9.55&&Tidal Pair\\
0879&14:01:30.8&$-$11:44:46&2400&0.44&0.57&9.72&45&\\
0883&14:03:15.8&$-$12:14:18&1800&0.64&0.65&10.10&89\\
0899&14:05:37.0&$-$12:23:38&1800&0.62&0.79&9.22&&Tidal Pair\\
0923&14:10:35.6&$-$12:25:05&2400&0.50&0.69&10.20&70\\
0944&14:16:30.6&$-$12:35:59&1200&0.62&0.62&10.00&62&Marginal\\
0954&14:20:29.7&$-$11:34:04&2400&0.39&0.67&17.30&115&Lens System\\
0961&14:24:16.7&$-$12:09:51&1800&0.49&0.74&11.10&74&\\
0974&14:28:59.7&$-$12:27:07&2400&0.52&0.63&10.90&85&\\
1007&14:46:08.4&$-$12:32:08&2400&0.46&0.53&13.10&94\\
1038&14:59:03.0&$-$12:51:08&1800&0.53&0.63&9.71&58\\
1050&15:01:43.4&$-$11:51:54&2400&0.51&0.68&9.34&34&Marginal\\
\enddata
\end{deluxetable}

\begin{deluxetable}{lcccrc}
\tablewidth{0pt}
\tablecaption{Candidate Arc Photometry}
\tiny
\tablehead{
\colhead{LCDCS}&\colhead{Lens Designation} &
\colhead{$g^\prime$}&\colhead{$r^\prime$}&\colhead{$g^\prime-r^\prime$}&\colhead{$R$}\\
\colhead{Cluster}}
\startdata
280&1&23.07&21.83&1.24&21.50\\
280&2&22.42&22.24&0.18&22.06\\
280&3&24.31&23.95&0.36&23.76\\
280&4&24.76&24.25&0.51&24.04\\
280&5&23.97&23.95&0.02&23.78\\
280&6&24.63&24.13&0.50&23.92\\
\\
486&1&20.74&20.07&0.67&19.81\\
486&1a&21.13&20.43&0.70&20.18\\
486&1b&22.08&21.40&0.68&21.14\\
\\
954&1&23.36&21.67&1.69&21.33\\
954&2&23.94&23.44&0.50&23.22\\
954&3&23.98&23.42&0.56&23.19\\
954&4&23.98&24.05&$-$0.07&23.91\\
\enddata
\end{deluxetable}

\end{document}